\documentclass{article}

\usepackage{arxiv}

\usepackage[utf8]{inputenc} % allow utf-8 input
\usepackage[T1]{fontenc}    % use 8-bit T1 fonts
\usepackage[pagebackref=true,breaklinks=true,colorlinks,bookmarks=false]{hyperref}
\usepackage{url}            % simple URL typesetting
\usepackage{booktabs}       % professional-quality tables
\usepackage{amsfonts}       % blackboard math symbols
\usepackage{nicefrac}       % compact symbols for 1/2, etc.
\usepackage{microtype}      % microtypography

\usepackage{graphicx}
\graphicspath{ {./images/} }

\usepackage{times}
\usepackage{epsfig}
\usepackage{graphicx}
\usepackage{amsmath}
\usepackage{amssymb}

\usepackage{cleveref}
\usepackage{algorithm}
\usepackage{algorithmic}
\usepackage{multirow}
\usepackage[caption=false]{subfig}
\usepackage{adjustbox}
\usepackage{caption}
\usepackage{xcolor}

\usepackage{xspace}

\makeatletter
\DeclareRobustCommand\onedot{\futurelet\@let@token\@onedot}
\def\@onedot{\ifx\@let@token.\else.\null\fi\xspace}

\def\eg{\emph{e.g}\onedot} 
\def\ie{\emph{i.e}\onedot} 
 
\def\etc{\emph{etc}\onedot} 
 
\def\etal{\emph{et al}\onedot}
\makeatother

\title{Nudge Attacks on Point-Cloud DNNs}

\author{
 Yiren Zhao, Ilia Shumailov, Robert Mullins and Ross Anderson \\
  Computer Laboratory\\
  University of Cambridge\\
  \texttt{name.surname@cl.cam.ac.uk} \\
}

\begin{document}
\maketitle
\begin{abstract}
The wide adaption of 3D point-cloud data in safety-critical applications such as autonomous driving makes adversarial samples a real threat. Existing adversarial attacks on point clouds achieve high success rates but modify a large number of points, which is usually difficult to do in real-life scenarios.
   In this paper, we explore a family of attacks that only perturb a few points of an input point cloud, and name them nudge attacks. 
   We demonstrate that nudge attacks can successfully flip the results of modern point-cloud DNNs. 
   We present two variants, gradient-based and decision-based, showing their effectiveness in white-box and grey-box scenarios.
   Our extensive experiments show nudge attacks are effective at generating both targeted and untargeted adversarial point clouds, by changing a few points or even a single point from the entire point-cloud input. We find that with a single point we can reliably thwart predictions in 12--80\% of cases, whereas 10 points allow us to further increase this to 37--95\%. Finally, we discuss the possible defenses against such attacks, and explore their limitations.
\end{abstract}

\section{Introduction}
Adversarial samples are now a major concern in machine learning.
An attacker can craft perturbations that are imperceptible to humans but cause DNN models to alter their results.
Existing adversarial sample generation has focused on natural language processing \cite{zhang2020adversarial,jia2017adversarial}, 2D images \cite{madry2017towards} and graph data \cite{zugner2018adversarial,dai2018adversarial}, among other targets.
Recent studies have revealed that 3D shapes represented as point clouds are also vulnerable to attacks \cite{xiang2019generating,liu2019extending}.
The limitation of existing point-cloud attacks is that they require a large number of points to be simultaneously perturbed. 
In reality, it is often feasible to modify large areas of a 2D image, for instance, a real-life spoofing attack can be implemented by projecting an image on a building in front of a car, or even by attaching colorful patches to street signs \cite{brown2017adversarial}. 
However, the physical limitations of 3D object sensors such as LiDAR (Light Detection and Ranging) make spoofing a large number of points unrealistic in practice \cite{cao2019adversarial}.
In addition, understanding adversarial point clouds created under extremely low point budgets, \eg by changing the position of a small number of points, gives insights to the sampling and feature extraction techniques used in current point-cloud DNNs.

\begin{figure}[!ht]
    \centering
    \subfloat[Original: monitor\label{1a}]{%
        \includegraphics[width=0.45\linewidth]{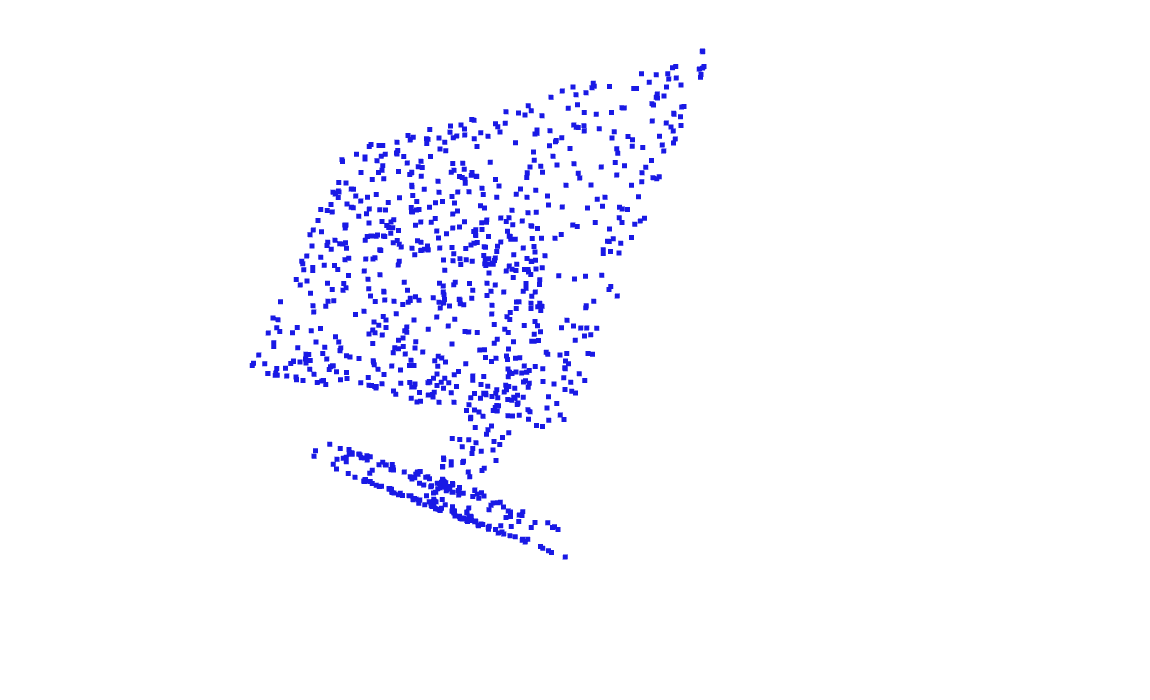}}
    \subfloat[1 point: chair\label{1b}]{%
        \includegraphics[width=0.45\linewidth]{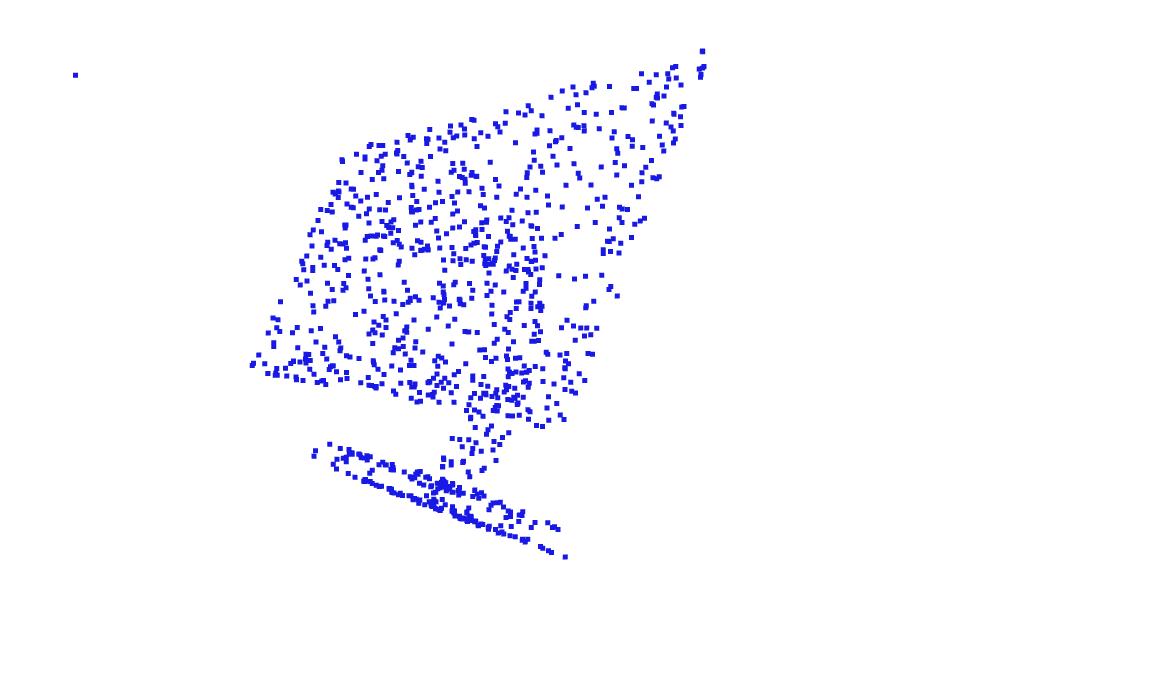}}
    \hfill
    \subfloat[10 points: chair\label{1c}]{%
        \includegraphics[width=0.45\linewidth]{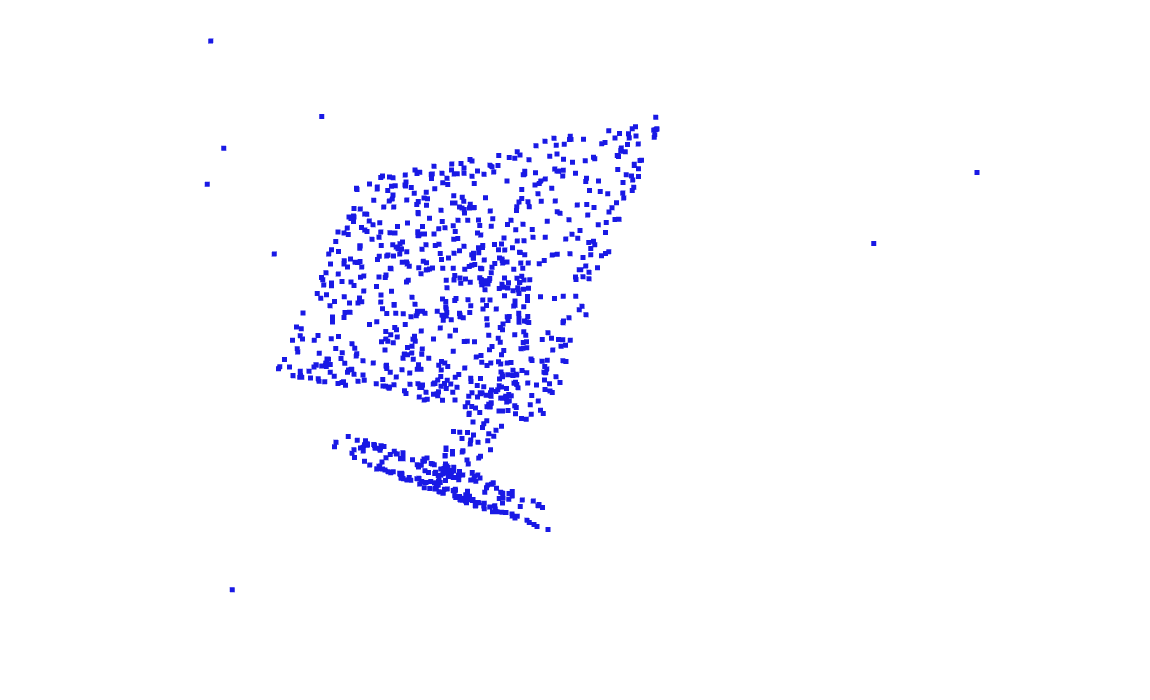}}
    \subfloat[150 points: table\label{1d}]{%
        \includegraphics[width=0.45\linewidth]{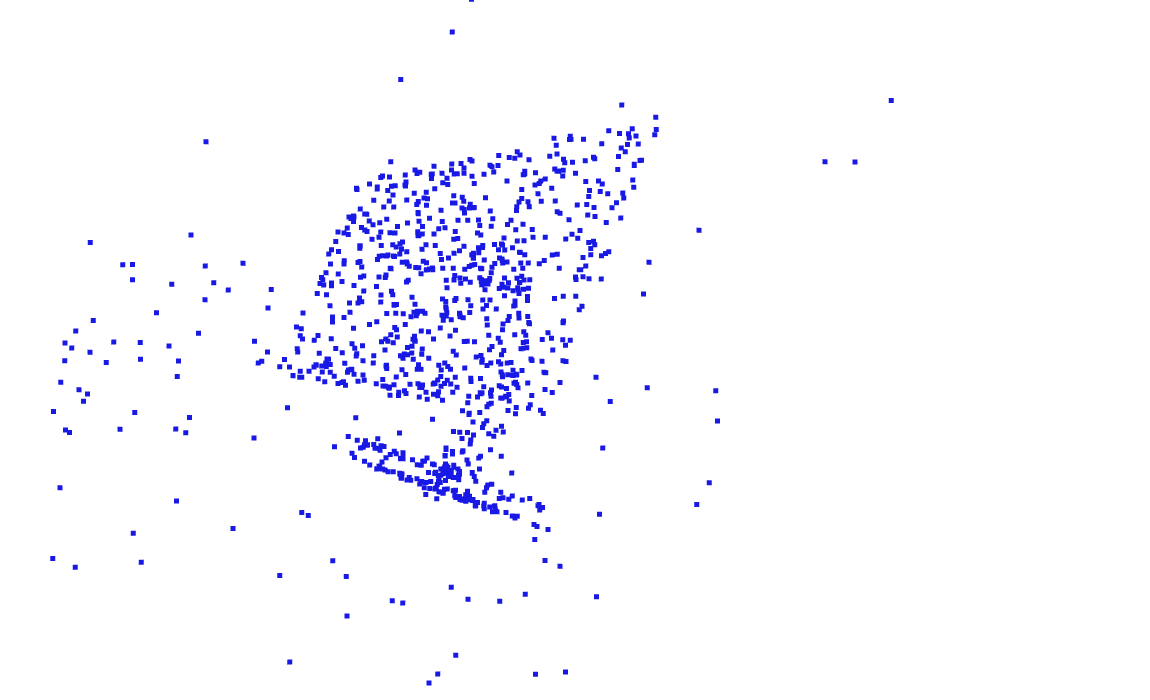}}
    \caption{Visualisation of untargeted nudge adversarial samples with different attack budgets on DGCNN, the DGCNN model makes a correct classification of the monitor from the ModelNet10 dataset.}
    \label{fig:visualisation} 
\end{figure}

In this work, we show that it is possible to fool point-cloud DNNs in both white-box and grey-box setups by only perturbing a few points. Furthermore, we show that such attacks work in both targeted and untargeted scenarios. We name them \textit{nudge attacks}, as they might be implemented by nudging a physical object near the target, by attaching small reflective objects, or equivalently by precision jamming of LiDAR using a laser.
As illustrated in \Cref{fig:visualisation}, perturbing a small number of points (1, 10 and 150) is enough to alter a point-cloud DNN's predictions from monitor to chair or table.
Restricting the number of points perturbed improves the practicality of launching such attacks on real systems, as it bridges the world of active and passive attacks on LiDAR sensing. Nudge attacks enable the attacker to thwart model predictions without having to have a large laser setup\footnote{To insert N points the attacker needs either N lasers or mirroring systems, and needs to solve an N-body problem of how to adjust them in both space and time.} -- indeed often inserting one point is enough.

The proposed nudge attack can find and then perturb vulnerable points from the input point cloud. This property helps the nudge attack to achieve success rates similar to existing methods with a much smaller attack budget.
Even the most limited variant, a single-point attack, shows a success rate of $36.54\%$ on changing the prediction results of a DGCNN  model~\cite{wang2019dynamic} trained for the popular ModelNet40 classification task.
Even with a simple defense, nudge attacks still achieve success rates that can create a real threat to deployed systems.
Adversarial samples generated using the nudge attack are much more portable than those made using previous methods~\cite{xiang2019generating}. Even when perturbing only a handful of points, the adversarial point clouds transfer well to other model architectures trained on the same dataset, making blind black-box attacks possible on point-cloud classifiers.

In this paper, we make the following contributions:
\begin{itemize}
    \item We introduce a family of attacks, the nudge attacks, for generating adversarial point clouds with limited attack budgets.
    \item We demostrate two variants, one gradient-based for white-box attacks and one based on an evolutionary algorithm for interactive grey-box attacks.
    From  extensive experiments, we demonstrate their effectiveness in both targeted and untargeted contexts.
    \item We show that the nudge attack has greater transferability across model architectures than existing attacks, creating a blind black-box threat to point-cloud DNNs.
\end{itemize}

\section{Background}
\textbf{Adversarial Examples.} Szegedy \etal~discovered that neural networks are vulnerable to carefully-crafted adversarial perturbations that are imperceptible to \cite{szegedy2013intriguing}.
There followed a rapid co-evolution of attack \cite{goodfellow2014explaining, kurakin2016adversarial, carlini2017towards} and defense \cite{madry2017towards, shumailov2020towards}.
Most of the gradient-based attacks (Fast Gradient Method \cite{kurakin2016adversarial}, Projected Gradient Descent \cite{madry2017towards}, \etc), focus on the loss function of the target neural network, producing adversarial samples that are very likely to fool it. 
Decision-based attacks \cite{brendel2017decision, su2019one} query the neural network interactively and require no gradient information, creating a real threat in black-box or grey-box setups.
A line of work that is particularly close to this paper is the previously proposed one-pixel attacks \cite{su2019one} on 2D images, where the attack focuses on scenarios where only one pixel (or a very few) can be modified in the input image.
They show that, using evolutionary algorithms in a grey-box setup and with access to the per-class probabilities only, one-pixel attacks can achieve a success rate of around $16\%$ on the ImageNet dataset.

\textbf{Adversarial Point Clouds.} Point clouds are sets of points that represent the shapes of 3D objects. They are widely used in applications in computer graphics and are raw outputs of many 3D sensing devices.
Qi \etal~first proposed to use sampling to reduce the high-dimensional unordered points to fixed-length feature vectors and process these features using Deep Neural Networks (DNNs), naming their network PointNet \cite{qi2017pointnet}.
Later, Wang \etal~ introduced the concept of edge convoltuion to extract shape information efficiently from point clouds, and constructed a network named Dynamic Graph CNN (DGCNN) \cite{wang2019dynamic}.

The exploration of adversarial samples then moved from 2D images to point clouds.
Liu \etal and Yang \etal~simply extend gradient-based adversarial attacks from 2D images to 3D point clouds \cite{liu2019extending, yang2019adversarial}.
Both Yang \etal~and Xiang \etal~tried to insert a set of carefully designed points into the original point clouds to fool classifiers \cite{yang2019adversarial,xiang2019generating}.
However, most of these point-cloud attacks perturb a lot of points, without considering the difficulty of actually doing such an attack in real life.

\textbf{Attacking Autonomous Driving.} 
Since modern autonomous driving relies heavily on point-cloud sensory data, a number of researchers have targeted this safety-critical application.
For instance, Cao \etal~ presented how real-world adversarial objects can evade LiDAR-based detection \cite{cao2019adversarialobjects} while Tu \etal~experimented with using 3D-printed adversarial objects on cars' rooftops to make them invisible \cite{tu2020physically}. Such attacks can in principle work on any point-cloud systems, including those that synthesize point clouds from multiple cameras. 
However, the direction of travel appears to be LiDAR, as such systems come down in price.
Thus another line of research, which is more closely related to our attack, focuses on 
remotely tampering with the LiDAR system using lasers \cite{petit2015remote,shin2017illusion,cao2019adversarial}.
Petit \etal~and Shin \etal~showed how to spoof LiDAR to introduce fake dots in their detection results \cite{petit2015remote,shin2017illusion}. They used a photodiode to delay the laser pulses fired by the victim LiDAR and sent an attack laser after certain delay to introduce fake points on the victim LiDAR. More generally, optical jamming systems used in electronic warfare tend to be be follower jammers which send a progressively delayed signal to `pull off' the attention of the sensor \cite{anderson2020}.
An attack of this kind can only generate a limited number of fake points, Cao \etal~reckoned that around 100 points can be spoofed, but only 60 points reliably \cite{cao2019adversarial}. 
This is an application that inspired our attack, bridging the world of passive and active attacks on LiDAR systems.
Attacks that employ fewer points pose a real threat in scenarios where the attacker has even more limited access to the system, \ie~due to SWaP (Size, Weight and Power) and cost constraints.
So while most spoofing attacks \cite{shin2017illusion,cao2019adversarial} focus on generating adversarial obstacles, adversarial samples appear to be the way of the future and enable more fine-grained manipulation of the system.

\section{Method}
\subsection{Gradient-based Nudge Attack}

The gradient-based nudge attacks we propose are white-box attacks that assume access to model parameters.
Consider a standard classification task with a data distribution $D$ that contains $d$-dimensional inputs $x \in \mathbb{R}^{d}$ and labels $y \in [k]$, where $k$ represents the number of classes. We assume a loss function $J(\theta, x, y)$, with model parameters $\theta$, and present the gradient-based attack algorithm in~\Cref{alg:fewpoints_grad}.

The attack first accumulates the gradients $n$ times to obtain $x_\delta$. Using the absolute difference $|x_\delta -x|$ between $x_\delta$ and $x$, and the $\text{top-k}$ function with the points budget $p$, the attack obtains a threshold value $\lambda$.
With the threshold, we can generate a mask $m$, and the binarised mask finds vulnerable regions in the input volume $x$. 
The final adversarial perturbation using the gradient $\epsilon \text{sign}(\bigtriangledown_{x_{\delta}}J(\theta, x_{\delta}, y)) \cdot m$ is similar to the PGD adversarial sample generation \cite{madry2017towards} but conditioned with a binarised mask $m$.
Note that \Cref{alg:fewpoints_grad} is used for generating untargeted adversarial examples. 
For a targeted attack, we generate perturbation towards a targeting class $y_t$, where we simply replace the mask generation and noise generation by 
$x_\delta = x_\delta - (\bigtriangledown_{x_{\delta}}J(\theta, x_{\delta}, y_t))$ and 
$x_\delta = x_\delta - \epsilon {sign}(\bigtriangledown_{x_{\delta}}J(\theta, x_{\delta}, y_t)) \cdot m$ respectively.
This is similar to other well-known targeted adversarial attacks in the literature. 

\begin{algorithm}
\caption{Gradient-based nudge attack algorithm}
\label{alg:fewpoints_grad}
\begin{algorithmic}
        \STATE {\bfseries Input:} $x$, $y$, $\epsilon$, $n$, $p$, $\theta$

        \STATE $x_\delta = x$
        \FOR{$i=0$ {\bfseries to} $n - 1$}
            \STATE $x_\delta$ = $x_\delta$ + ($\bigtriangledown_{x_{\delta}}J(\theta, x_{\delta}, y)$)
        \ENDFOR
        \STATE $\lambda$ = top-k($|x_\delta-x|$, $p$)
        \STATE $m$ = binarise($x_\delta \geq \lambda$)
        \STATE $x_\delta = x$
        \FOR{$i=0$ {\bfseries to} $n - 1$}
            \STATE $x_\delta$ = $x_\delta$ + $\epsilon$ sign($\bigtriangledown_{x_{\delta}}J(\theta, x_{\delta}, y)$) $\cdot$ $m$
        \ENDFOR
\end{algorithmic}
\end{algorithm}

\subsection{Evolution-based Nudge Attack}

Inspired by the one-pixel attack on CNNs, we designed a nudge attack using Differential Evolution (DE)~\cite{su2019one}. DE is an optimisation algorithm from a family of evolutionary algorithms operating over populations of candidate points~\cite{storn1997differentialevolution}. DE is known to produce high-quality solutions and avoid local minima that are problematic for gradient-based methods. 

DE does not assume differentiability and requires no access to the network internals apart from the per-class probabilities of the targeting model. The algorithm relies on feedback from a fitness function that assesses how well the candidate samples perform. All best-performing candidates are used to generate a new improved pool, and a random mutation is applied for diversity. The interactive grey-box nature of DE makes it a powerful and realistic tool for attacks in the real world -- often the attacker may have little or even no knowledge of system internals. Indeed, current literature has developed a range of evolutionary algorithm-based attacks targeting different properties of ML systems~\cite{nguyen2015deep,xu2016automatically,shumailov2020sponge}.

\section{Evaluation} 

\subsection{Datasets and Settings}
We evaluate the nudge attack on multiple datasets, including the ModelNet10 and ModelNet40 datasets~\cite{wu20153d} for classification and the Stanford 3D Large-Scale Indoor Spaces (S3DIS) dataset for semantic segmentation \cite{armeni20163d}.

On the ModelNet10 datasets, there are 10 object classes, 3991 training samples and 908 test samples. The ModelNet40 dataset has 9843 training objects and 2468 test objects.
We sample the data to 1024 points; each point is 3-dimensional with $x$, $y$ and $z$ spatial coordinates, the same as PointNet~\cite{qi2017pointnet}. The S3DIS dataset has 6 sampled indoor areas with 271 rooms, with in total 13 categories for the semantic segmentation. We follow the convention on this dataset to use Area 5 as the test data, training on Areas 1, 2, 3, 4 and 6~\cite{qi2017pointnet}. We sample 4096 points for each scanned scene.
For the reported attack performance, we use 256 point clouds from the test set and report their average performance unless specified otherwise. 
We take the network implementations from Pytorch Geometric \cite{fey2019fast} and
report the details of attack hyperparameters and network training in our Appendix. In what follows, `Adv Accuracy' refers to the accuracy of adversarial point clouds and `Success Rates' refer to the percentage of adversarial samples that successfully changed the classification results.

\begin{figure*}[!htp]
    \includegraphics[clip,width=\linewidth]{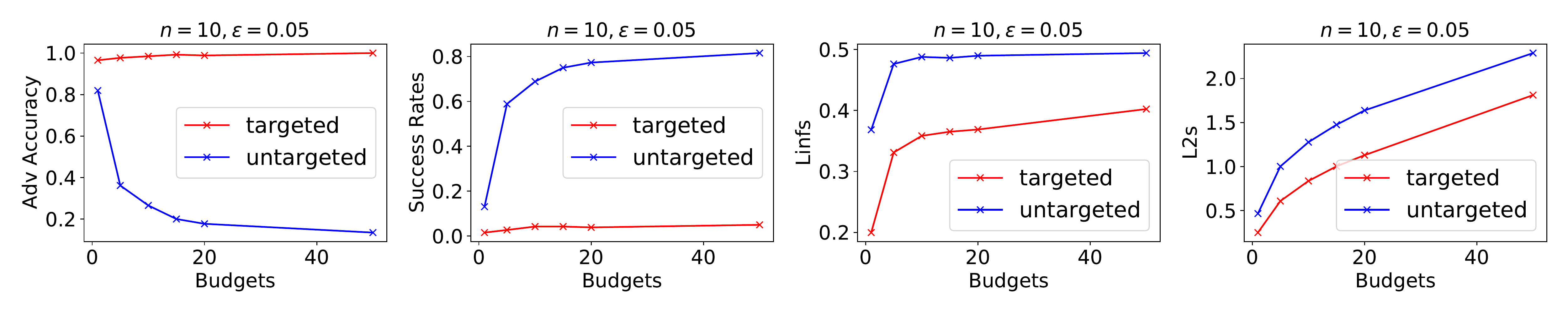}
    \includegraphics[clip,width=\linewidth]{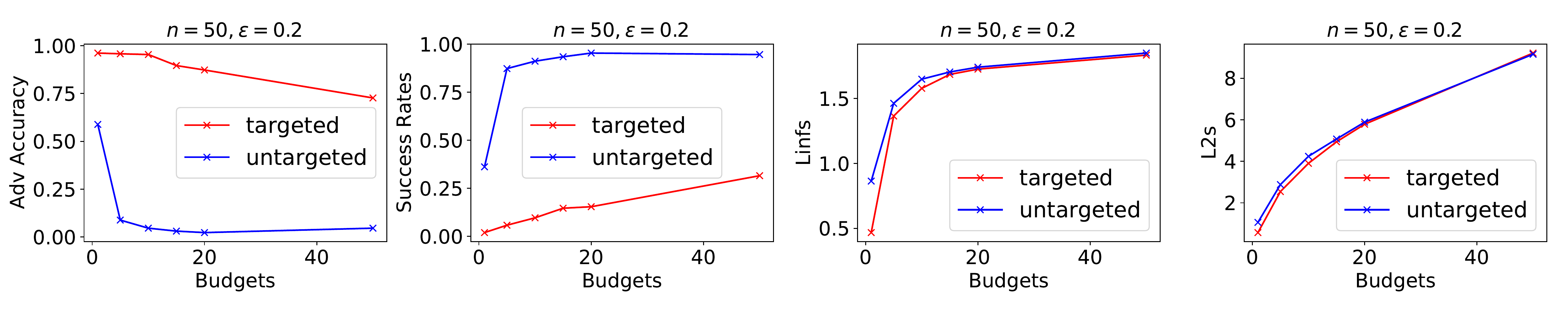}
    \includegraphics[clip,width=\linewidth]{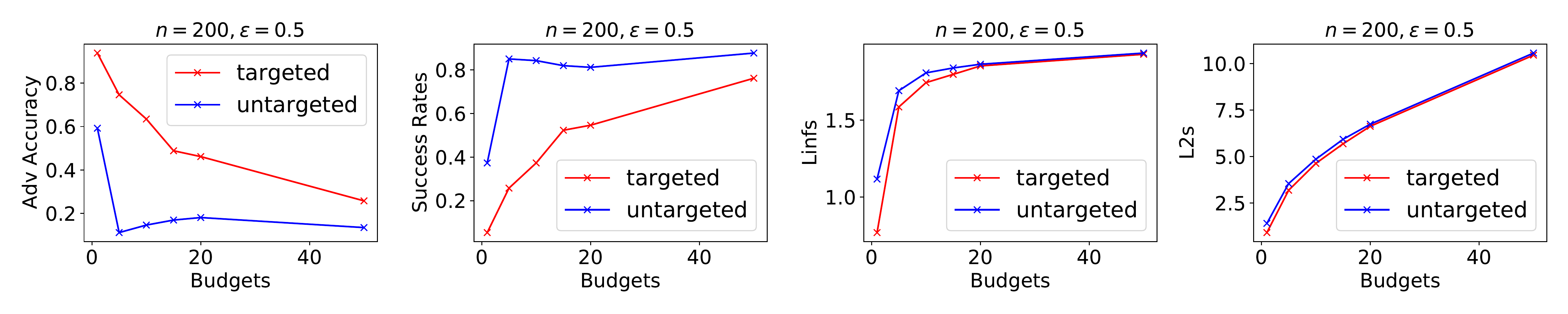}
    \caption{
        Targeted and untargeted gradient-based white-box nudge attack with a weak, a moderately strong and a strong adversary. The original DGCNN model achieves a $93.72\%$ accuracy on the ModelNet10 classification task. The horizontal axis shows the number of points allowed to be edited, while the vertical axis shows adversarial accuracy, success rates, $L_{\infty}$ and $L_{2}$ norms of 256 input point clouds from the evaluation set. Details are discussed in \Cref{sec:eval:grad_attack}.}
    \label{fig:grad_attack_dgcnn}
\end{figure*}

\begin{figure*}[!h]
    \begin{center}
    \includegraphics[width=.7\linewidth]{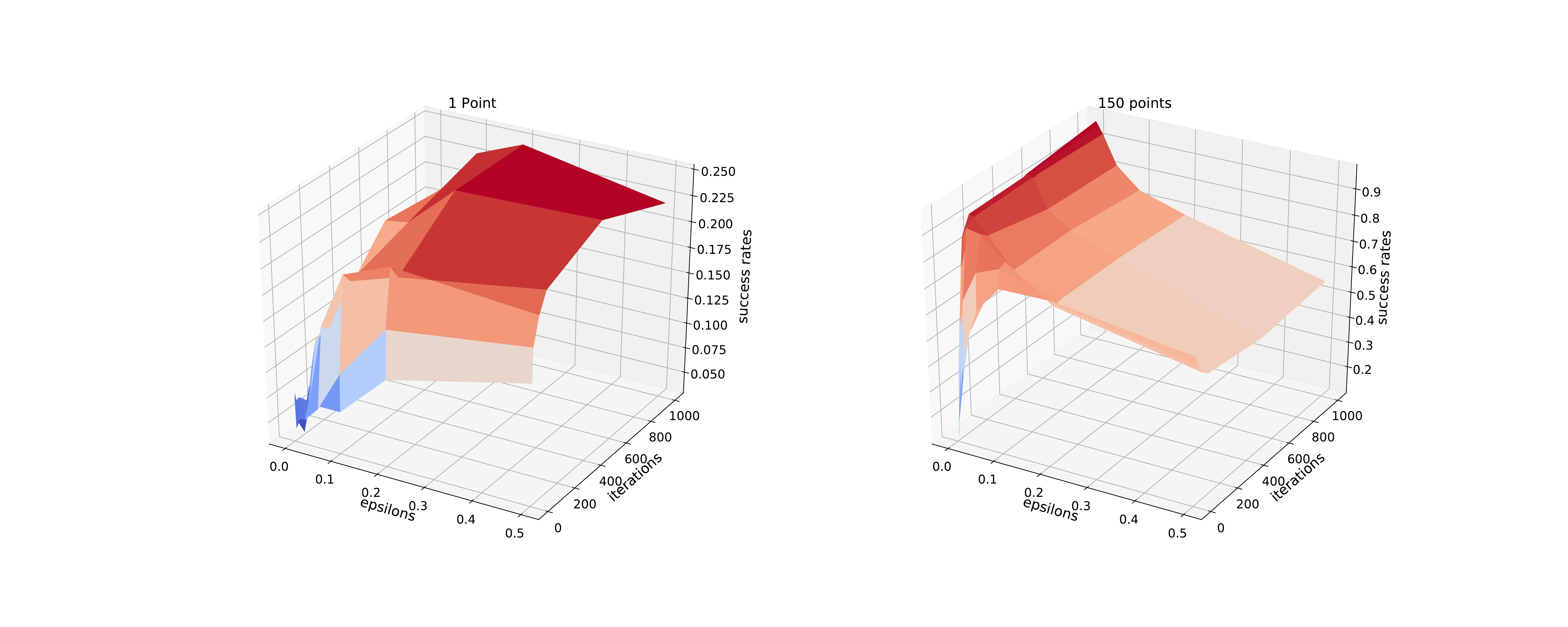}
    \end{center}
    \caption{Performance of one-point and 150-point nudge attacks with different per-iteration perturbation sizes ($\epsilon$) and number of iterations ($n$).
    Details are discussed in \Cref{sec:eval:grad_attack}.
    }
    \label{fig:grad_attack_param}
\end{figure*}

\begin{table*}[h!]
    \caption{
        Gradient-based nudge attack performance on the S3DIS dataset, a semantic segmentation task.
        We use Nudge-U and Nudge-T to represent untargeted and targeted attacks respectively. Details are discussed in \Cref{sec:eval:grad_attack}.
    }
    \label{tab:s3dis}
    \vskip 0.15in
    \begin{center}
    \begin{small}
    \begin{sc}
    \begin{tabular}{cc|cccc}
    \toprule
    &
    & \multicolumn{2}{c}{PointNet}
    & \multicolumn{2}{c}{DGCNN}
    \\
    Attack Method
    & Points Edited
    & Accuracy
    & Success Rates 
    & Accuracy
    & Success Rates 
    \\
    \midrule
    Natural
    &
    & 63.72
    & -
    & 86.90
    & -
    \\ 
    \midrule
    \multirow{3}{*}{Nudge-U}
    & 1
    & 58.04
    & 26.82
    & 63.08
    & 32.69
    \\
    & 10
    & 48.11
    & 36.82
    & 42.23
    & 54.73
    \\
    & 150
    & 48.07
    & 37.03
    & 38.86
    & 58.13
    \\
    \midrule
    \multirow{3}{*}{Nudge-T}
    & 1
    & 59.50
    & 25.03
    & 74.86
    & 20.11
    \\
    & 10
    & 49.71  
    & 34.91
    & 58.19 
    & 39.02
    \\
    & 150
    & 48.52 
    & 36.41
    & 52.98 
    & 44.42 \\
    \bottomrule
    \end{tabular}
    \end{sc}
    \end{small}
    \end{center}
    \vskip -0.1in
\end{table*}

\subsection{Gradient-based Nudge Attack}
\label{sec:eval:grad_attack}
The gradient-based nudge attack is white-box, because it assumes access to the model parameters and architecture.
The baseline model is DGCNN \cite{wang2019dynamic}, achieving an accuracy of $93.72\%$ on the evaluation dataset.
We first demonstrate the effectiveness of a gradient-based nudge attack on three adversaries when given different point budgets.
The three adversaries are:
\begin{itemize}
    \item A weak adversary with $n=10, \epsilon=0.05$
    \item A moderately strong adversary with $n=50, \epsilon=0.2$
    \item A strong adversary with $n=200, \epsilon=0.5$
\end{itemize}

$n$ and $\epsilon$ represent the number of iterations and the per-iteration perturbation size respectively (\Cref{alg:fewpoints_grad}).
We evaluate how nudge attackers perform targeted and untargeted attacks with these three different configurations in \Cref{fig:grad_attack_dgcnn}.
For targeted attacks, we randomly pick a class that is not the original class as the attack target; in an untargeted attack, the aim is simply make the network generate prediction results different from the original ones.

\Cref{fig:grad_attack_dgcnn} shows several general trends:
\begin{itemize}
    \item Unsurprisingly, untargeted attacks show greater performance than targeted ones.
    \item More points give better attack performance, but performance is only marginally better when using more than 20 points for untargeted attacks.
\end{itemize}
For untargeted attacks, we found a weak adversary with a moderate point budget ($>15$) can consistently achieve high attack success rates ($>70\%$).
However, targeted attacks are harder, and typically require a strong adversary. This is a justification for our name, a `nudge' attack; if you can nudge a shooter's elbow just before they fire, you can make them miss the target, but it takes more strength or skill to nudge them into hitting a different target instead.
And indeed we see that, when point budgets are smaller, strong adversaries can provide better attack performance.

By testing different adversaries, we realised the hyperparameter choices of the gradient-based nudge attack can greatly influence the attack performance.
We then apply a grid-search for $\epsilon$ and $n$ (details can be found in Appendix).
% $\epsilon = [0.001, 0.002, 0.005, 0.01, 0.02, 0.05, 0.1, 0.2, 0.5]$ and $n = [5, 10, 50, 100, 500, 1000]$, where $\epsilon$ and $n$ determine the perturbation size and number of iterations respectively.
This time, we look at the ModelNet10 classification task using a PointNet model with an accuracy of $89.99\%$ on clean data.
\Cref{fig:grad_attack_param} shows the success rates surface of untargeted attacks with different epsilon values and number of iterations when given a single-point and a 150-point budget. We use these two representative cases to support the observations made below, but in general, we see these trends across different point budgets (more results are in the Appendix).
The effect of different hyperparameter combinations of the gradient-based nudge attack is shown in \Cref{fig:grad_attack_param}, and we observed that:
\begin{itemize}
    \item Given a small point budget, a large $\epsilon$ and lot of iterations will give better attack success rates.
    \item Given a large point budget, a small $\epsilon$ and lot of iterations gives better attack success rates.
\end{itemize}

To further demonstrate the effectiveness of nudge attacks, we show their performance on a different task and dataset in \Cref{tab:s3dis}.
\Cref{tab:s3dis} shows how nudge attacks operate on the semantic segmentation task on the S3DIS dataset.
We use the attack parameters found from the PointNet ModelNet10 grid search.
The S3DIS dataset is significantly larger and takes a longer time to run. Due to limited facilities, we report results averaged across 128 samples from the evaluation dataset.
The results in \Cref{tab:s3dis} suggest that nudge attacks scale well to larger datasets and different point-cloud tasks.
Our results reveal that PointNet is more resistant than DGCNN when the perturbation budget is large. We hypothesise that this is because of the sampling strategy of PointNet which focuses more on local information. In addition, the natural accuracy of PointNet is lower, so it is harder to decrease the adversarial accuracy further. 
We will further provide a more throughout comparison of different styles of nudge attacks on more datasets and model architectures in \Cref{tab:all}.

\subsection{Evolution-based Nudge Attack}
\label{sec:eval:de}

\begin{figure*}[!h]
    \begin{center}
    \includegraphics[width=\linewidth]{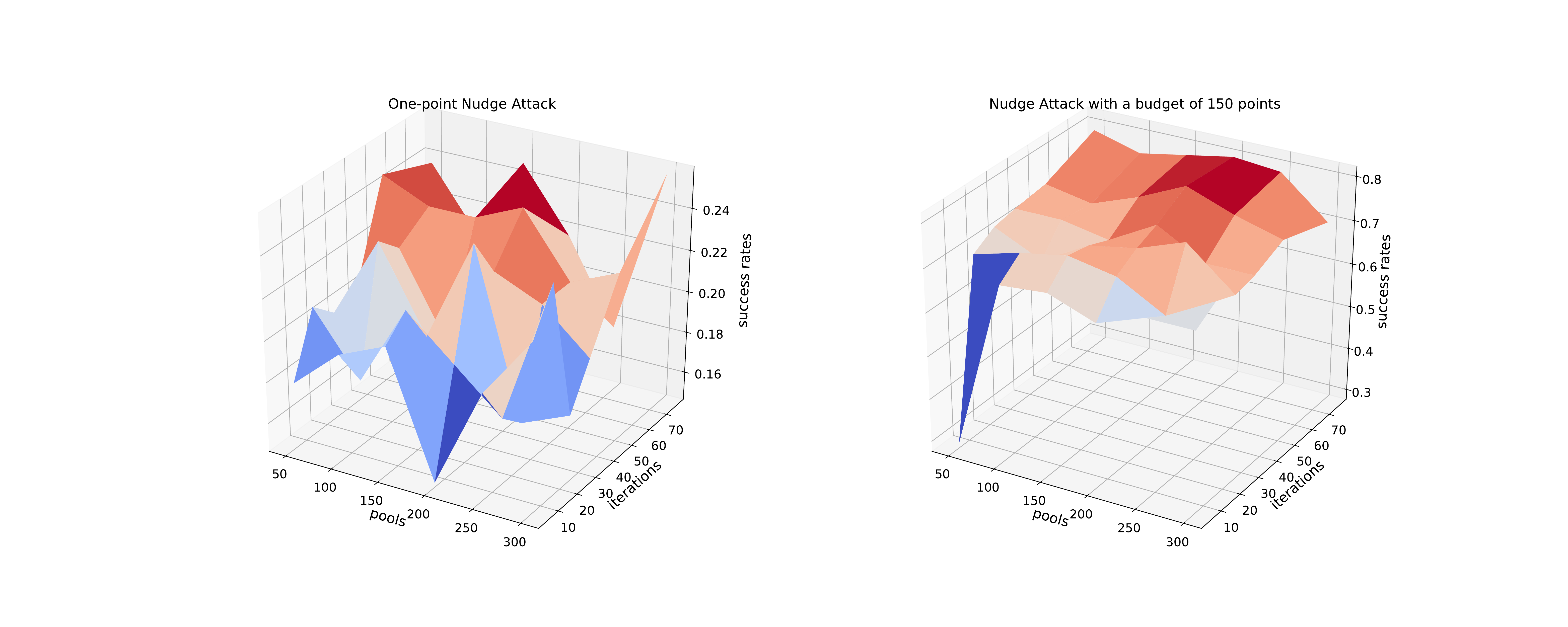}
    \end{center}
    \caption{Performance of an evolution-based one-point and 150-point nudge attacks with different pool sizes and number of iterations.
    Details are discussed in \Cref{sec:eval:de}.
    }
    \label{fig:de_attack_param}
\end{figure*}

\Cref{fig:de_attack_param} shows the performance of the differential evolution-based nudge attack. 
The increased pool size leads to an improvement in attack effectiveness for both one-point and nudge variants. 
Interestingly we observe that the attack converges using only a few iterations. 
More iterations will lead to better performance, but the improvement is marginal. In what follows we report the results of Evolution-based nudge attack with 10 iterations and 100 pool size.

\begin{table*}[h!]
    \caption{
        Comparison with existing attacks on the PointNet architecture trained on the ModelNet40 dataset. Pointwise Gradient represents a wide range of attacks that use gradients on all points directly \cite{xiang2019generating,yang2019adversarial, liu2019extending}. For nudge attacks, we sample 256 point clouds from the test set and report their average readings in the table. Details are discussed in \Cref{sec:eval:comp}.
    }
    \label{tab:comparison}
    \vskip 0.15in
    \begin{center}
    \begin{small}
    \begin{sc}
    \begin{tabular}{c|ccccc}
    \toprule
    Method                          & Adv Accuracy      & Success Rate      & $l_2$     &  Points Edited \\
    \midrule
    Random Noise                    & $82\%$            & $12.01\%$            & 1.64      & 1 \\
    Random Noise                    & $78\%$            & $18.02\%$            & 3.42      & 10 \\
    Random Noise                    & $79\%$            & $18.34\%$            & 5.19      & 150 \\
    % Random Noise                    & $20.01\%$         & $79.84\%$         & 3.55       & 100 \\
    Point-Attach \cite{yang2019adversarial}                 & $75.12\%$         &  -                & -         & 100 \\
    Point-Detach \cite{yang2019adversarial}                 & $67.58\%$         &  -                & -         & 200 \\
    Pointwise Gradient              & $4.23\%$          & $95.77\%$         & 8.18      & 1003 \\
    % Adversarial Clusters \cite{xiang2019generating}         & -                 &  $99.3\%$         & -         & 768 \\
    % Adversarial Objects \cite{xiang2019generating}          & -                 &  $97.3\%$         & -         & 192 \\
    \midrule
    1-point nudge                       & $68.01\%$ &  $25.00\%$ & $1.68$    & 1 \\
    10-point nudge                       & $33.46\%$ &  $61.65\%$ & $2.13$    & 10 \\
    100-point nudge                       & $1.83\%$ &   $93.75\%$ & $3.42$    & 100 \\
    150-point nudge                       & $0.37\%$  &  $97.42\%$ & $5.29$    & 150 \\
    200-point nudge                       & $0.73\%$  &  $98.16\%$ & $8.74$    & 200 \\
    \bottomrule
    % Adversarial Objects \cite{}     & PointNet      & -                 &  $97.3\%$         & -         & 192
    % \midrule
    % AGNN$^*$                & float  & $83.6 \pm 0.3\%$  & $200.0$KB   & $73.8 \pm 0.7 \%$  & $2840.0$KB     & $79.7 \pm 0.4 \%$ & $280.0$KB\\
    % PDNAS                   & float  & $84.5 \pm 0.6 \%$ &$1534.9$KB   & $73.5 \pm 0.3 \%$  & $3824.9$KB     & $79.7 \pm 0.6 \%$ & $812.7$KB\\
    % \midrule
    % Ours & $\mathbf{84.5 \pm 0.6 \%}$ & $73.5 \pm 0.3 \%$ & $\mathbf{79.7 \pm 0.6 \%}$\\
    \end{tabular}
    \end{sc}
    \end{small}
    \end{center}
    \vskip -0.1in
\end{table*}

\begin{table*}[h!]
    \caption{
        Attack performance on different datasets and models.
        Acc and Rates represent adversarial and success rates of the adversarial samples.
        Budget means the number of points we can edit in an adversarial point cloud.
        WN means gradient-based white-box nudge attack, and DE means its DE-based black-box alternative.
        We use -U and -T to represent untargeted and targeted attacks respectively. Details discussed in \Cref{sec:eval:comp}.
    }
    \label{tab:all}
    \vskip 0.15in
    \begin{center}
    \begin{small}
    \begin{sc}
    % \begin{adjustbox}{width=\textwidth}
    % \begin{tabular}{cc|cccccccccccc}
    \begin{tabular}{cc|cccccccc}
    \toprule
    &
    & \multicolumn{4}{c}{ModelNet10}      
    & \multicolumn{4}{c}{ModelNet40} 
    % & \multicolumn{4}{c}{S3DIS} 
    \\
    & 
    & \multicolumn{2}{c}{PointNet}
    & \multicolumn{2}{c}{DGCNN}
    & \multicolumn{2}{c}{PointNet}
    & \multicolumn{2}{c}{DGCNN}
    % & \multicolumn{2}{c}{PointNet}
    % & \multicolumn{2}{c}{DGCNN}
    \\
    Attack
    & Budget
    & Acc
    & Rates
    & Acc
    & Rates
    & Acc
    & Rates
    & Acc
    & Rates
    % & Acc
    % & Rates
    % & Acc
    % & Rates
    \\ 
    \midrule
    Natural
    & -
    & 89.88
    & -
    & 93.72
    & -
    & 89.18
    & -
    & 92.38
    & -
    % & 63.72
    % & -
    % & 86.90
    % & -
    \\ 
    \midrule
    \multirow{3}{*}{WN-U}
    & 1
    & 78.07
    & 12.30
    & 60.00
    & 35.38
    & 68.01
    & 25.00
    & 58.84
    & 36.54
    % & 48.04
    % & 36.82
    % & 63.08
    % & 32.69
    \\
    & 10
    & 55.76
    & 37.69
    & 3.46
    & 92.31
    & 33.46
    & 61.65
    & 7.31
    & 90.00
    % & 48.11
    % & 36.82
    % & 42.23
    % & 54.73
    \\
    & 150
    & 11.15
    & 82.31
    & 0.77
    & 95.77
    & 0.37
    & 97.42
    & 0.31
    & 97.31
    % & 48.07
    % & 37.03
    % & 38.86
    % & 58.13
    \\
    \midrule
    \multirow{3}{*}{DE-U}
    & 1
    % Pointnet model10
    & 73.26
    & 20.79
    % DGCNN model10
    & 14.85
    & 80.19
    % Pointnet model40
    & 62.37
    & 32.67
    % DGCNN model40
    & 28.71
    & 66.33
    \\
    & 10
    % Pointnet model10
    & 57.42
    & 38.61
    % DGCNN model10
    & 0.99
    & 95.04
    % Pointnet model40
    & 33.66
    & 63.36
    % DGCNN model40
    & 7.92
    & 87.12
    \\
    & 150
    & 28.71
    & 67.32
    
    & 0.00
    & 100.00
    
    & 9.90
    & 89.10
    
    & 0.99
    & 97.02
    \\
    \midrule
    \multirow{3}{*}{WN-T}
    & 1
    & 90.38
    & 5.77
    & 82.69  
    & 12.69
    & 83.08 
    & 11.54
    & 91.53
    & 3.85
    % & 47.97 
    % & 36.89
    % & 77.54
    % & 17.61
    \\
    & 10
    & 78.85
    & 16.15
    & 33.08 
    & 66.54
    & 60.38
    & 35.77
    & 28.85
    & 69.62
    % & 48.13  
    % & 36.76
    % & 66.60 
    % & 30.01
    \\
    & 150
    & 71.92
    & 23.85
    & 18.85
    & 81.15
    & 54.23
    & 43.08
    & 15.00 
    & 85.00
    % & 47.95 
    % & 37.66
    % & 56.39 
    % & 41.00 
    \\
    \midrule
    \multirow{3}{*}{DE-T}
    & 1
    % Pointnet model10
    & 82.49
    & 10.89
    % Dgcnn model10
    & 59.40
    & 40.59
    % Pointnet model40
    & 76.26
    & 19.84
    % Dgcnn model40
    & 76.23
    & 17.82
    \\
    & 10
    % Pointnet model10
    & 80.93
    & 14.00
    % Dgcnn model10
    & 32.67
    & 63.36
    % Pointnet model40
    & 59.14
    & 37.74
    % Dgcnn model40
    & 50.49
    & 48.51
    \\
    & 150
    % Pointnet model10
    & 64.20
    & 33.85
    % Dgcnn model10
    & 26.73
    & 73.26
    % Pointnet model40
    & 31.90
    & 65.75
    % Dgcnn model40
    & 18.81
    & 79.20
    \\
    \bottomrule
    \end{tabular}
    % \end{adjustbox}
    \end{sc}
    \end{small}
    \end{center}
    \vskip -0.1in
\end{table*}

\subsection{Comparison to Existing Methods}
\label{sec:eval:comp}
\Cref{tab:comparison} shows how gradient-based nudge attacks, in a white-box setup, compare with existing attacks on the ModelNet40 dataset. 
The baseline PointNet achieves a $89.88\%$ clean accuracy.
We use the grid search results from \Cref{sec:eval:grad_attack} to configure the nudge attacks, whose parameters can be found in the Appendix. 

The random noise attack serves as a baseline; we match the $l_2$ norm of random noise attacks to nudge attacks for an easier comparison.
In the random noise attack, we randomly sample points from the point cloud and send scaled noise to the randomly selected points.
The pointwise gradient attack is a representative of several proposed attacks that directly use gradients to construct adversarial point clouds \cite{liu2019extending,xiang2019generating,yang2019adversarial}. We attack the same target network (PointNet)~\cite{qi2017pointnet} as previous work to help ensure a fair comparison. In \Cref{tab:comparison}, some methods, such as point-attach and point-detach \cite{xiang2019generating}, add or delete points from the original point clouds; so their total number of points in the adversarial point clouds might differ from the original ones.
This might allow a potential defense that simply rejects samples with extra points.
In contrast, our method only edits points in the existing point cloud and does not add or delete points. 
% It is also worth mentioning that Yang \etal~claimed their Adversarial Clusters and Adversarial Objects as targeted attacks, however, their evaluation only focused on perturbing classification results to one particular class.
% Also, attackers under this assumption in have to obtain training data information of a class.
All reported attacks in \Cref{tab:comparison} are untargeted.
Apart from the random noise baseline, all other reported attacks are white-box.

The results in \Cref{tab:comparison} suggest that nudge attacks show similar performance to existing attacks \cite{xiang2019generating,yang2019adversarial} given an attack budget of 150 or 200 points, while the compared attacks use significantly more of points.
More importantly, we demonstrate that with a single point budget, the attack success rate can be as high as $25.00\%$, making it a feasible simple attack in practice.
In addition, our nudge attacks can find vulnerable points in the input point cloud. When compared to random noise attacks with the same $l_2$ and point budget, nudge attacks show significantly better results as they hit the vulnerable points.

In \Cref{tab:all}, we show the attack performance on DGCNN and PointNet of two different datasets.
We display both the adversarial accuracy (Acc) and success rates (Rates) of both targeted and untargeted nudge attacks.
We grid search the attack parameters on PointNet attacking ModelNet10, and use the best hyperparameter combination we find for all other model-dataset pairs.
In \Cref{tab:all}, WN means gradient-based white-box nudge attack, and DE is the DE-based grey-box or black-box alternative.
We use -T and -U after the attack to indicate whether the attack is targeted or not.
Budget represents the number of points allowed to perturb and we show both the adversarial accuracy (ACC) and success rates (Rates) in the table as well.

\Cref{tab:all} shows a general trend that a larger point budget makes both targeted and untargeted attacks easier.
It is worth noticing that, in most cases, DGCNN is more vulnerable than PointNet in gradient-based attacks, although it is a more complex model.
This phenomenon is can be seen on the ModelNet10 comparison in \Cref{tab:comparison}, where PointNet is more robust in both targeted and untargeted attacks.
This observation is opposite to the one made by Mardy \etal on Convolutional Neural Networks \cite{madry2017towards}, where they found that complex models have better adversarial robustness.
We suggest the robustness of point cloud models depends on the per-layer sampling strategy used by the network. 
For instance, the global k-nearest neighbors sampling used by DGCNN makes the network more vulnerable to adversarial attacks since it receives perturbations globally. 
The PointNet architecture, in contrast, samples local points, reducing the the amount of noise injected per point.
Because of the above reason, DE-based nudge attacks, since it does not need gradient information, show better performance on PointNet compared to the gradient-based alternative. 

\begin{table*}[h!]
    \caption{
        Attack transferability across model architectures on ModelNet40. The adversarial sample is generated for a PointNet model (the source model). PointNet* is a PointNet model trained using a different initialisation. Details are discussed in \Cref{sec:eval:transfer}.
    }
    \label{tab:transfer}
    \vskip 0.15in
    \begin{center}
    \begin{small}
    \begin{sc}
    \begin{tabular}{cc|cccc}
    \toprule
    Attack Method
    & Target Model
    & Success Rates 
    & Points Edited
    \\ 
    \midrule
    Adversarial Cluster \cite{xiang2019generating}
    & DGCNN
    & 16.9\%
    & 768
    \\ 
    Adversarial Objects \cite{xiang2019generating}
    & DGCNN
    & 16.5\%
    & 192
    \\
    Adversarial Cluster \cite{xiang2019generating}
    & PointNet* 
    & 48.3\%
    & 768
    \\ 
    Adversarial Objects \cite{xiang2019generating}
    & PointNet* 
    & 39.2\%
    & 192
    \\ 
    \midrule
    1-point nudge
    & DGCNN
    & 33.5\%
    & 1
    \\ 
    % Few-Points 
    % & DGCNN
    % & 60.7\%
    % & 10
    % \\ 
    150-point nudge 
    & DGCNN
    & 69.5\%
    & 150
    \\ 
    1-point nudge
    & PointNet* 
    & 19.9\%
    & 1
    \\ 
    150-point nudge
    & PointNet* 
    & 52.2\%
    & 150
    \\ 

    % One-Points
    % & SplineCNN
    % & 8.0\%
    % & 1
    % \\ 
    % Few-Points
    % & SplineCNN
    % & 16.0\%
    % & 10
    % \\ 
    \bottomrule
    \end{tabular}
    \end{sc}
    \end{small}
    \end{center}
    \vskip -0.1in
\end{table*}

\subsection{Transferability Across Model Architectures}
\label{sec:eval:transfer}
The transferability of adversarial samples across model architectures is important, since it can enables a blind black-box attack.
An attacker with no knowledge of the target model can blindly use adversarial samples constructed from another model.
An attack with high transferability may work here, and also scale to different real-life scenarios.

Xiang \etal~previously studied the transferability of adversarial samples, but found that their adversarial point clouds did not transfer well even when the attack is untargeted.
Their Adversarial Cluster and Adversarial Objects, while adding a large number of points, only show around $16\%$ success rates when transferred to DGCNN \cite{xiang2019generating}.
We provide a detailed comparison of transferability in \Cref{tab:transfer}.
The source model for which we prepared our adversarial samples is a PointNet model, while the target models are DGCNN and a PointNet trained from a different initialisation (PointNet*).
The results reveal that nudge attacks provide better transferability compared to Xiang \etal, and their performance is worse than the white-box case displayed in \Cref{sec:eval:grad_attack}.

\subsection{Attacking Past a Blind Defense}
\label{sec:eval:defense}
\begin{figure}[!htp]
    \includegraphics[clip,width=\linewidth]{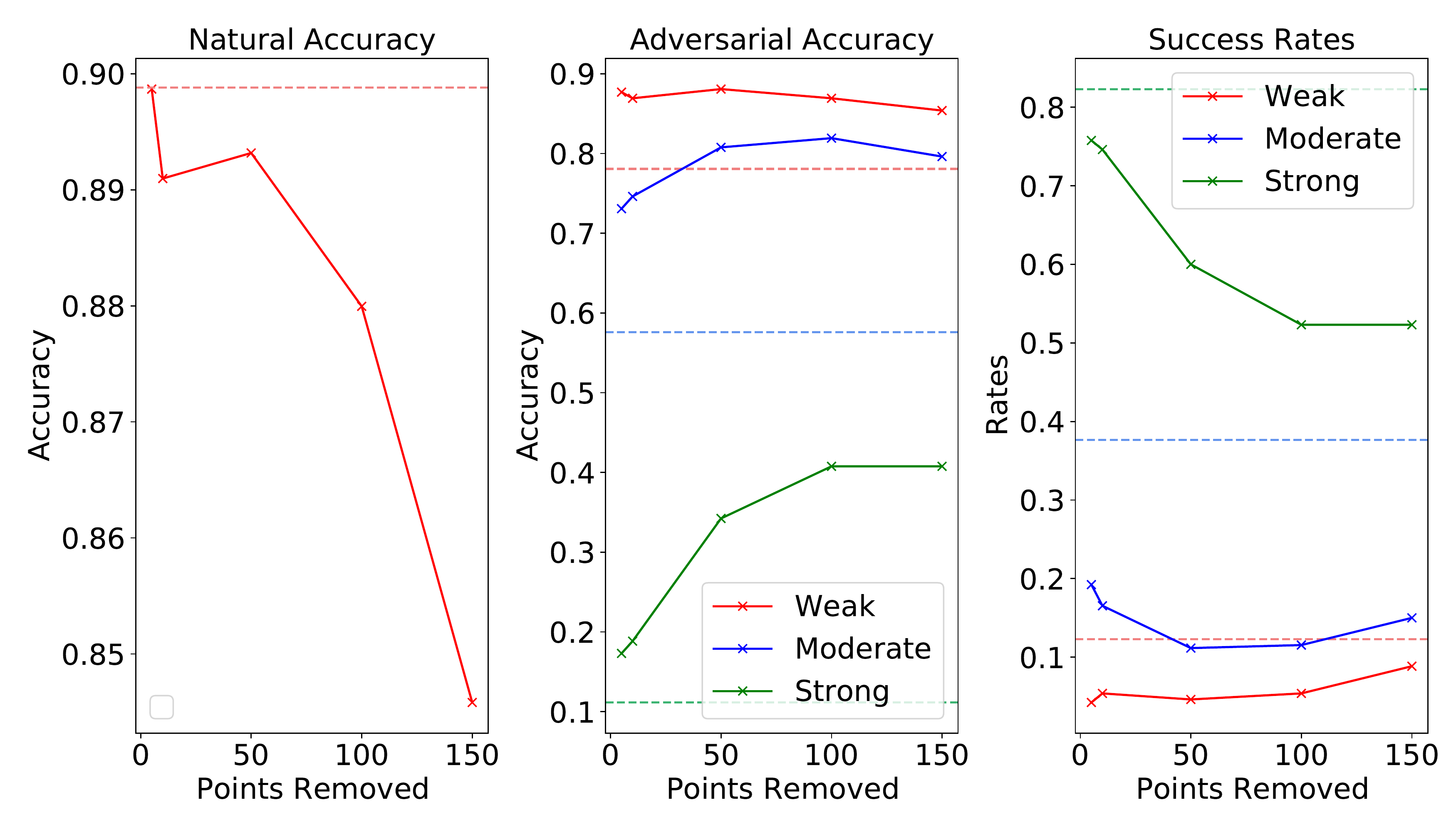}
    \caption{Performance of nudge attacks with a point-removal defense. Weak, moderate and strong attackers mount nudge attacks with increasing capability; dashed horizontal lines show performance without any defense. Details are discussed in \Cref{sec:eval:defense}.}
    \label{fig:defense}
\end{figure}

A natural defense to adversarial  attacks on point clouds is to remove points that are far from the original clouds. So we first normalise all points to make them centralise at the origin and then filter out the top-k furthest points from the origin. We call this a point-removal defense, and show the detailed algorithm in Appendix. It is our strawman blind defense in that it assumes nothing about the likely attack.
\Cref{fig:defense} shows how nudge attacks on PointNet perform against such a defense. 

We consider weak (1 point), moderate (10 points) and strong (150 point) attackers, and compare their performance in the absence of the defense in dashed horizontal lines. The point-removal defense does indeed provide a certain baseline of protection; the dashed lines always show higher success rates in \Cref{fig:defense}. 

Could such defenses be improved to block nudge attacks more effectively? We invite the community to find such defenses, but we can see four reasons why they are unlikely to block all attacks. 
First, as the point removal budget increases, the success rates start to flatten, providing more limited protection.
Second, point removal imposes a growing cost in model accuracy.
Third, nudge attacks still show relatively large success rates (around $20\%$ and $70\%$) for moderate and strong attackers in the face of point removal.
Finally, our evaluation only considers the ModelNet10 dataset. Naturally collected point clouds in the wild, such as raw LiDAR inputs, are normally noisy and this may allow nudge attacks to hide adversarial points better.
The point-removal defense might be improved if the defender knows enough about the attacker's statistics to optimise the removal strategy. However, this takes us out of the scope of blind defense as it requires some prior knowledge of the attack and is beyond the scope of this paper.

\section{Conclusion}
In this paper, we introduced a family of attacks, the nudge attacks.
We discussed two variants, gradient-based attacks for a white-box setup and evolutionary attacks for the interactive grey-box case.
Our attacks require the jammer to change significantly fewer points than previous attacks, bringing them within the range of operational feasibility. As an example, attackers may now have the  capability of manipulating point cloud DNNs while attacking LiDAR-based autonomous driving systems. 
What's more, our nudge attacks are highly transferable across model architectures, making them a realistic candidate for a blind black-box attack.
Finally, we show that a blind point-removal defense is not enough to protect against them.

\clearpage

{\small
\bibliographystyle{ieee_fullname}
\bibliography{references}
}

\appendix

\section{Network Training Configurations}
In \Cref{tab:networks}, we show the training configurations, including learning rates, number of epochs and the final accuracy, of all networks with different datasets that we have used in the paper.
We took all of the model implementations from Pytorch Geometric \cite{fey2019fast}.

\begin{table}[h!]
    \caption{
        Network Training Configurations.
    }
    \label{tab:networks}
    \vskip 0.15in
    \begin{center}
    \begin{small}
    \begin{sc}
    % \begin{adjustbox}{width=\textwidth}
    % \begin{tabular}{cc|cccccccccccc}
    \begin{tabular}{ccccc}
    \toprule
    Datasets 
    & Networks
    & Epochs 
    & lr 
    & Accuracy
    \\
    \midrule
    ModelNet10
    & PointNet 
    & 150 
    & 0.1
    & 89.88\%
    \\
    ModelNet10
    & DGCNN
    & 150 
    & 0.1
    & 93.72\%
    \\
    ModelNet40
    & PointNet 
    & 150 
    & 0.1
    & 89.18\%
    \\
    ModelNet40
    & DGCNN 
    & 150 
    & 0.1
    & 92.38\%
    \\
    S3DIS
    & PointNet 
    & 150 
    & 0.01
    & 64.72 \%
    \\
    S3DIS
    & DGCNN 
    & 150 
    & 0.01
    & 86.90\%
    \\
    \bottomrule
    \end{tabular}
    % \end{adjustbox}
    \end{sc}
    \end{small}
    \end{center}
    \vskip -0.1in
\end{table}

\section{Differential Evolution Algorithm}

The evolutionary algorithm used in nudge attacks is the differential evolution algorithm, which is the same as one-pixel attack \cite{su2019one}.
We present the full algorithm in \Cref{alg:de_algo}, the idea is to have pools of candidates and keep mixing the best performing candidates to generate new pools.
\begin{algorithm}
\caption{Differential evolution}
\label{alg:de_algo}
\begin{algorithmic}
        \STATE {\bfseries Input:} $x$, $y$, Pool size $N$, Attack budget $n$, fitness function $f$, crossover parameter $CR$, iterations $iters$, mutation factor $m$

        \STATE $A : N\times n = random$ (init of pool)
        \FOR{$i=0$ {\bfseries to} $iters$}
            \STATE $A_{new} : N\times n = zero$
            \FOR{$j=0$ {\bfseries in} $N$}
                \STATE $r1 = \text{random-from}(A)$
                \STATE $r2 = \text{random-from}(A)$
                \STATE ${t} = A_{best} + m * (r_1 - r_2)$
                
                \STATE $mask = CR > \text{random-of-size}(n)$
                \STATE $a_{temp} = t * mask + A[j] * (1-mask)$
                
                \IF{$f(a_{temp}) > f(A[j])$}
                    \STATE $A_{new}[j] = a_{temp}$
                \ELSE
                    \STATE $A_{new}[j] = A[j]$
                \ENDIF
            \ENDFOR
            \STATE $A = A_{new}$
        \ENDFOR
\end{algorithmic}
\end{algorithm}

\section{Attack Parameters}
We found attack parameters of the gradient-based nudge attack using a grid search on the PointNet \cite{qi2017pointnet} model on the ModelNet10 dataset.
The grid search has the search space of 
\begin{itemize}
    \item  $\epsilon = [0.001,0.002,0.005,0.01,0.02,0.05,0.1,0.2,0.5]$ 
    \item $n = [5,10,50,100,500,1000]$.
\end{itemize}

$\epsilon$ and $n$ are the the per-iteration perturbation size and number of iterations respectively, the detailed definitions can be found in Algorithm 1.
We then use this attack configuration for gradient-based nudge attacks for all the other networks on different datasets.
\begin{table}[h!]
    \caption{
        Searched Attack Parameters for gradient-based Few points attack used in Table 1, 2, 3 and Figure 5.
    }
    \label{tab:attackparam}
    \vskip 0.15in
    \begin{center}
    \begin{small}
    \begin{sc}
    % \begin{adjustbox}{width=\textwidth}
    % \begin{tabular}{cc|cccccccccccc}
    \begin{tabular}{cccc}
    \toprule
    Point Budget 
    & $\epsilon$ 
    & $n$ 
    & mode
    \\
    \midrule
    1
    & 0.5 
    & 500 
    & untargeted
    \\
    10
    & 0.1
    & 1000 
    & untargeted
    \\
    150
    & 0.05
    & 1000 
    & untargeted
    \\
    1
    & 0.02 
    & 500 
    & targeted
    \\
    10
    & 0.5
    & 100 
    & targeted
    \\
    150
    & 0.5
    & 100 
    & targeted
    \\
    \bottomrule
    \end{tabular}
    % \end{adjustbox}
    \end{sc}
    \end{small}
    \end{center}
    \vskip -0.1in
\end{table}

\begin{figure*}[!h]
    \includegraphics[clip,width=\linewidth]{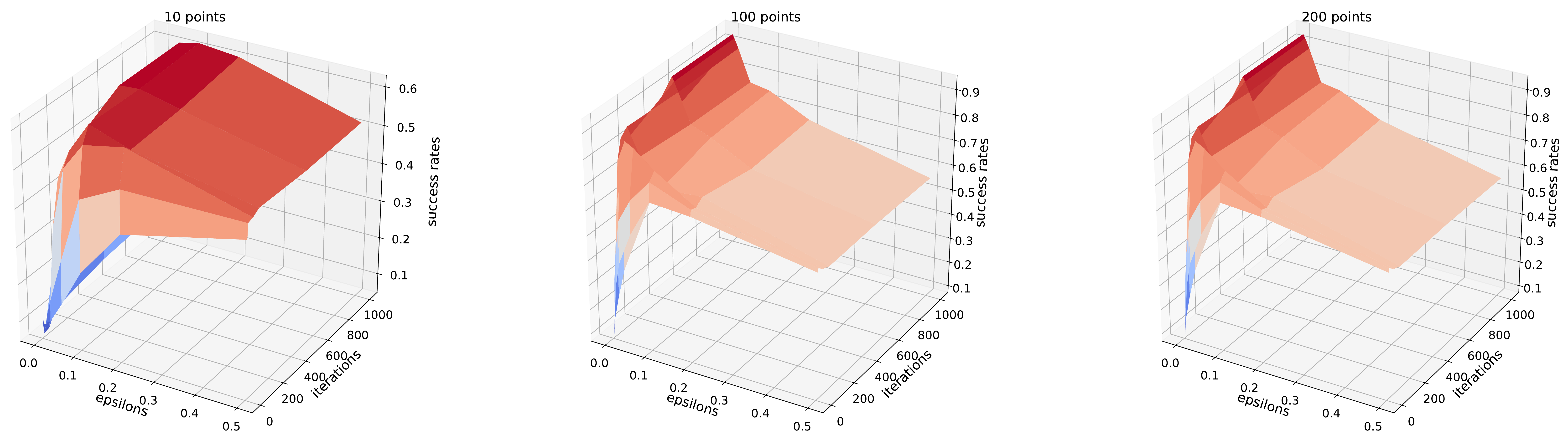}
    \caption{Success rates surfaces when applied gradient-based nudge attacks with different attack budgets.}
    \label{fig:attack_budgets}
\end{figure*}
\section{Different Attack Budgets}
In \Cref{fig:attack_budgets}, we show the success rates surfaces of different attack budgets.
This confirms again with our previous observation that:
\begin{itemize}
    \item Given a small point budget, a large $\epsilon$ and lot of iterations will give better attack success rates.
    \item Given a large point budget, a small $\epsilon$ and lot of iterations gives better attack success rates.
\end{itemize}

Our additional results in \Cref{fig:attack_budgets} show that these observations hold true for various attack budgets.

\end{document}